# The cage effect in systems of hard spheres

W. van Megen[1] and H. J. Schöpe[2]

[1]*Department of Applied Physics, Royal Melbourne Institute of Technology, Melbourne, Victoria 3000, Australia*

[2]*Institut für Angewandte Physik, Universität Tübingen, Auf der Morgenstelle 10, 72076 Tübingen, Germany*

ABSTRACT

The cage effect is generally invoked when discussing the delay in the decay of time correlation functions of dense fluids. In an attempt to examine the role of caging more closely we consider the spread of the displacement distributions of Brownian particles. These distributions are necessarily biased by the presence of neighbouring particles. Accommodation of this bias by those neighbours conserves the displacement distribution locally and presents a collective mechanism for exploring configuration space that is more efficient than the intrinsic Brownian motion. Caging of some particles incurs, through the impost of global conservation of the displacement distribution, a delayed, non-local collective process. This non-locality compromises the efficiency with which configuration space is explored. Both collective mechanisms incur delay or stretching of time correlation functions, in particular the particle number and flux densities. This paper identifies and distinguishes these mechanisms in existing data from experiments and computer simulations on systems of particles with hard sphere interactions.

**Keywords**; hard spheres, colloids, classical fluids, collective modes, glass transition, freezing transition.

# I. INTRODUCTION

The "cage effect" describes the transient dynamical confinement of a molecule by its neighbouring molecules. Associated, collective phonon excitations, or energy "hotspots", promote chemical reactions in solutions[1] in one context and, in another, activate processes that lead to restoration of ergodicity in amorphous solids[2]. Aside from being a source of these irreversible processes, caging is also considered in fluids generally as the underlying cause for the delay in the decay, relative to that of an exponential, of time correlation functions as seen, for example, by the sub-diffusive region in the evolution of the particle's mean-squared displacement (MSD) or by stretching in the decay of the correlation function of the particle number density [3-7]. The questions one might raise by this picture are, how long do those structures or clusters, comprising particles confined by their nearest and possibly successive layers of neighbours, persist? And while they persist, is their movement accompanied by a backflow, so that density may be conserved[8, 9]? Finally, how does such dynamical heterogeneity, comprising caged and more freely moving particles, dissipate? In other words, how does a particle "escape" its neighbour cage?

Rahman[10], in one of the earliest molecular dynamics (MD) computer simulations, answers the last question as follows; "the displacement of a particle…is the same as the direction in which, on the average, the instantaneous configuration of its neighbours shows a characteristic elongation…… the particle at the centre (of the cage) takes advantage of the fluctuation to move in the 'easy' direction afforded to it by the fluctuation". This observation applies to liquid argon above the melting temperature. Just over 30 years later Donati et al.[11] in an MD study of a supercooled binary mixture of particles interacting with the Lennard-Jones potential found "that it is much more probable for a mobile particle to move in the direction of another mobile particle than any other direction." This picture of the cooperation by which particles follow each other over distances beyond the inter-particle spacing will tend to intensify the "wings" of the (average) particle displacement distribution (PDD). This in turn, as found in the cases just mentioned, is manifested by positive deviations of the PDD from a Gaussian [12, 13]. So, on the face of it, it seems that this cooperative cage "escape" mechanism features in dense fluids whether they are in thermodynamic equilibrium or supercooled. At the same time, the fact that in the supercooled case non-Gaussian effects are much larger and more sensitive to temperature [13] suggests there is some other mechanism, not present above the melting temperature. Phonon assisted "hopping" is perhaps most commonly considered to be that mechanism[14-18]. This would seem to be supported by microscopic studies which certainly give the appearance of particles hopping between relatively immobile clusters of particles[19-21]. But then consistency in this regard, and in the manner by which the decay of time correlation functions is similarly delayed, in simulations of particles subject to Newtonian and diffusive dynamics [22-24] suggests that, at some level, large displacements of cooperating particles may be a consequence of excluded volume effects rather than activated by (undamped) phonons.

We have just identified, however loosely, caging, cooperative cage escape and phonon-assisted hopping; any of which could underpin non-Gaussian displacement statistics and non-diffusive decays of the structure as expressed, for instance, by time correlation functions of the particle number or current densities. However, there is a lack of clarity in precisely when these mechanisms emerge and how they compete. With these identifications the questions raised above might be rephrased as; What non-Fickian, non-Gaussian processes are common to thermodynamically stable and metastable fluids? What mechanisms, if any, differentiate these states? Do caged particles trigger a cooperative escape mechanism, like that suggested by Rahman, or phonon-activated jumps? And how can these mechanisms be identified in experimental and computational data?

Since we are essentially concerned with the dynamical consequences of packing or excluded volume among the particles, the discussions in this paper will refer extensively to the system of hard spheres. This system is completely specified by the packing fraction, $\phi$; key values are those for which fluid, $\phi_f=0.494$, and crystal, $\phi_m=0.545$, phases coexist [25, 26], random close packing, $\phi_R \approx 0.64$, and $\phi_g \approx 0.57$, where a glass transition (GT) has been identified [27].

*Background.* To justify a re-examination of the role of the cage effect we consider the mean squared displacements, $<\Delta r^2(\tau)>$, for suspensions of hard spheres obtained previously by dynamic light scattering (DLS)[28, 29]. Fig. 1a shows some typical results. The largest deviation from diffusion, or maximum stretching, occurs where the exponent $\gamma$ in the growth of the MSD, $<\Delta r^2(\tau)> \sim \tau^\gamma$, is smallest. The location of this point is the given by the delay time, $\tau_m$, and the root-mean-squared (RMS) displacement, $R_m$; $R_m^2=<\Delta r^2(\tau_m)>$. The stretching exponent, $\gamma$, and $\tau_m$, read from these data are shown in Fig. 1b for packing fractions from just above zero to the GT. $R_m$ along with the average distance, $R_c=[\phi_R/\phi]^{1/3}-1$, between particle surfaces in are shown in Fig. 1c.

For packing fractions between $\phi_f$ and $\phi_g$ the above results, as well as those presented in the remainder of this paper, apply to the metastable suspension. The integrity of this state is able to be maintained long enough, by a small spreads in particle radii[30], to measure (stationary) time correlation functions. At the same time, the polydispersities are still small enough ($\lesssim 10\%$) that the suspensions ultimately crystallise so that the packing fractions of the observed coexisting fluid and crystal phases can be mapped onto those of the perfect hard sphere system and the samples' effective hard-sphere packing fractions, $\phi$, defined[26, 31].

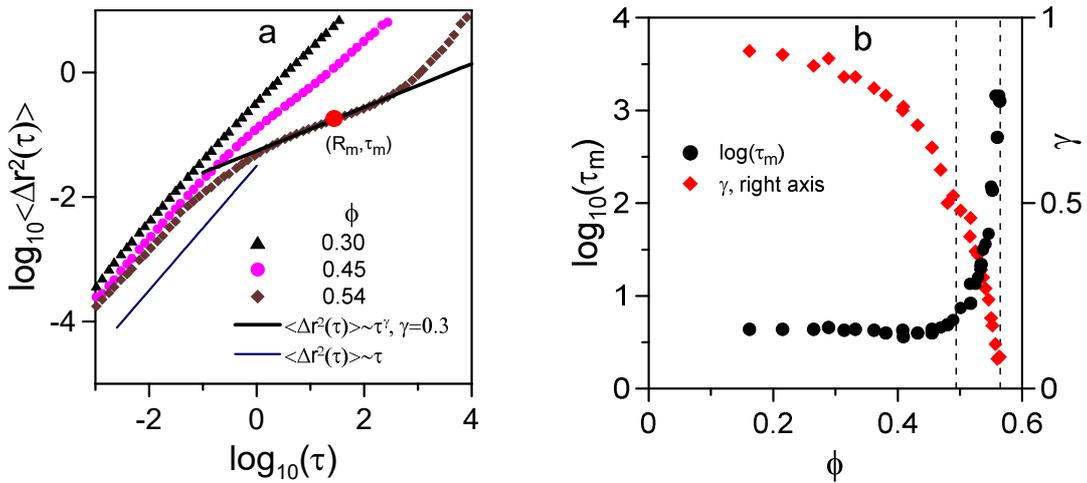

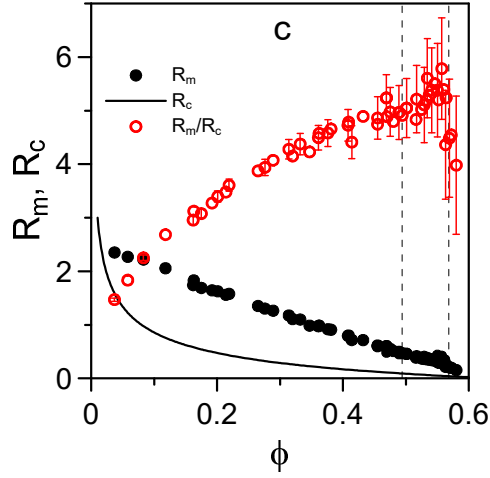

FIG. 1. Results for hard sphere system. (a) Double logarithm plots of the mean-squared displacements versus delay time at packing fractions indicated; (b) Delay time, $\tau_m$, and stretching index, $\gamma$. (c) The RMS displacement, $R_m$, and average surface to surface distance, $R_c$. The thin, dashed vertical lines indicate the freezing and glass transition packing fractions, $\phi_f=0.494$ and $\phi_g\approx0.57$, respectively. In these and following results distances are expressed in units of the relevant particle radii, a, and delay times in units of the characteristic Brownian time, $\tau_b=a^2/6D_o$; $D_o$ is the diffusion coefficient of a freely diffusing particle. See text for further details and ref. [28, 29, 32] for similar complementary results.

Were stretching due to cage confinement alone one might expect $R_m$ (Fig. 1c) to be bounded by $R_c$, at least approximately. However, as is evident from the ratio, $R_m/R_c$, this is far from the case; near the freezing packing fraction, $\phi_f=0.494$, for instance, $R_m/R_c\approx5$. One might also expect the "caging time", $\tau_m$, to increase uniformly with $\phi$. However, in Fig. 1b one sees that $\tau_m$ shows no systematic variation with $\phi$ up to around $\phi_f$ and then it increases rather sharply. Despite the errors in $R_m/R_c$ for higher values of $\phi$ (>$\phi_f$), it's difficult to rationalise these results in Fig.1 generally in terms of the cage effect alone; one suspects *some other collective process*, one that involves accommodating cage fluctuations, may also play a role.

The remainder of this paper is structured as follows; In Sec. II.A we give a rudimentary description of the theory and we note, in particular, that delay in dephasing of the particle number density field can result from either constraint on the magnitude or direction of the particles' displacements. In Sec. II.B we consider a system of Brownian particles in one dimension. We build on this simple model to lead to criteria that allow us to identify these constraints by two mechanisms; caging and correlation of cage fluctuations. In Sec. III.A these mechanisms are then identified in extant experimental and computational data. Sec. III.B considers briefly further dynamical consequences of caging. In these Sections most of the discussion concerns suspensions of hard-sphere like particles, a system where sound is effectively damped by the suspending liquid. So, the collective dynamics consequent upon excluded volume effects are able to be identified for all length scales. However, as we discuss briefly, for hard spheres subject to Newtonian dynamics, as simulated by MD, such identification is limited to just those conditions for which sound is overdamped. Conclusions are presented in Sec. IV.

## II. THEORY

### A. Time correlation functions

The basic dynamical property of concern is the intermediate scattering function (ISF) or the (normalized) time auto-correlation function,

$$f(q,\tau)=\langle\rho(q,0)\rho^\dagger(q,\tau)\rangle/\langle|\rho(q)|^2\rangle \qquad (1)$$

of the $q^{th}$ spatial Fourier component of the particle number density,

$$\rho(q,t) = \sum_{k=1}^{N} \exp[-i\mathbf{q}\cdot\mathbf{r}_k(t)], \qquad (2)$$

where $\tau$ is the delay time, $\mathbf{r}_k(t)$ the position of particle k at time t, and "†" denotes the complex conjugate. The ISF is normalised by the static structure factor, $\langle|\rho(q)|^2\rangle=S(q)$. Note that delay in the decay, or de-phasing, of the density field, $\rho(q,t)$, as manifested by stretching of the ISF, can result from constraint on the magnitude of the particles' movements to their neighbour cages (say $|\Delta\mathbf{r}_k|<R_c$) and/or persistence of movements beyond the cage in the direction of the propagation vector $\mathbf{q}$ ($\mathbf{q}\cdot\Delta\mathbf{r}_k/q>R_c$). Accordingly, we tentatively define these sources of delay as caging (C) and escape (E). The second, being anisotropic, is exposed more directly by the (normalised) autocorrelation function[33, 34],

$$C(q,\tau)=q^2\langle j(q,0)j^\dagger(q,\tau)\rangle/\langle|\rho(q)|^2\rangle=-d^2f(q,\tau)/d\tau^2, \qquad (3)$$

of the longitudinal particle current density

$$j(q,t) = \sum_{k=1}^{N} \hat{\mathbf{q}}\cdot\mathbf{v}_k(t)\exp[-i\mathbf{q}\cdot\mathbf{r}_k(t)], \qquad (4)$$

where $\hat{\mathbf{q}}=\mathbf{q}/q$, and $\mathbf{v}_k(t)$ is the velocity of particle k at time t. In the results discussed below $C(q,\tau)$, which we'll refer to as simply the current auto-correlation function (CAF), has been obtained by numerical differentiation of $f(q,\tau)$ [35, 36].

For a colloidal suspension the fastest processes that can be detected by conventional DLS are considered to be diffusive[4]. This allows a change of variable, $\tau^*=q^2D(q)\tau$, where the coefficient $D(q)=d/d\tau[f(q,\tau\to"0")/q^2]$ characterises the initial decay of the ISF and "0" refers to the lower detection limit (typically of order $10^{-5}$s.). We refer to the result,

$$C^*(q,\tau^*)=-d^2f(q,\tau^*)/d\tau^{*2}, \qquad (5)$$

as the scaled current correlation function (SCAF). The self ISF,

$$f_s(q,\tau) = \langle\exp[i\mathbf{q}\cdot\Delta\mathbf{r}(\tau)]\rangle, \qquad (6)$$

where $\Delta\mathbf{r}(\tau)$ is the displacement of a "tagged" particle executed in a time interval $\tau$, is a complementary dynamical property of the system. Its cumulant expansion is [37]

$$f_s(q,\tau)=-\frac{q^2}{6}\langle\Delta r^2(\tau)\rangle - \frac{1}{2}\left(\frac{q^2}{6}\right)^2\langle\Delta r^2(\tau)\rangle^2\alpha(\tau) + \cdots, \qquad (7)$$

where $\langle\Delta r^2(\tau)\rangle$ is the particle's mean-squared displacement (MSD) and

$$\alpha(\tau) = \frac{3}{5}(\langle\Delta r^4(\tau)\rangle - \langle\Delta r^2(\tau)\rangle^2)/\langle\Delta r^2(\tau)\rangle^2 \qquad (8)$$

is the leading order deviation of the PDD from Gaussian, generally referred to as the non-Gaussian parameter. For the Gaussian PDD the (even) moments are related as follows;

$$\langle r^{2n}\rangle=C_n\langle r^2\rangle^n, \qquad C_n=1\times 3\times 5\times\ldots\ldots(2n+1)/3^n. \qquad (9)$$

In the limit of extreme dilution particles diffuse independently of each other and the above time correlation functions, given in Eq. (1), (3) and (6), decay exponentially with delay time; specifically,

$f(q,\tau)=f_s(q,\tau)=\exp[-D_o q^2 \tau]$. (10)

where $D_o$ is the free particle diffusion constant; $D_o=<\Delta r^2(\tau)>/6\tau$. At finite $\phi$ any delay dephasing of the single particle density field ($\exp[i\mathbf{q}.\mathbf{\Delta r}]$), by either mechanism C or E, is manifested by stretching of the MSD; ie, $<\Delta r^2(\tau)>\sim\tau^\gamma$ with $\gamma<1$ (see Fig. 1a). The minimum values of the exponent $\gamma$ (Fig. 1b) expose maximum impacts of these mechanisms.

## B. Caging and escape in a system of Brownian particles.

To facilitate identification of mechanisms C and E in time correlation functions we begin by considering a system of Brownian hard spheres. This simplification eliminates processes activated by undamped phonons from the discussion, for now. Consider this in one dimension; a line of particles with packing fraction $\phi=1/(1+R_c)$, where $R_c$ is the average gap between neighbouring particles (Fig. 2).

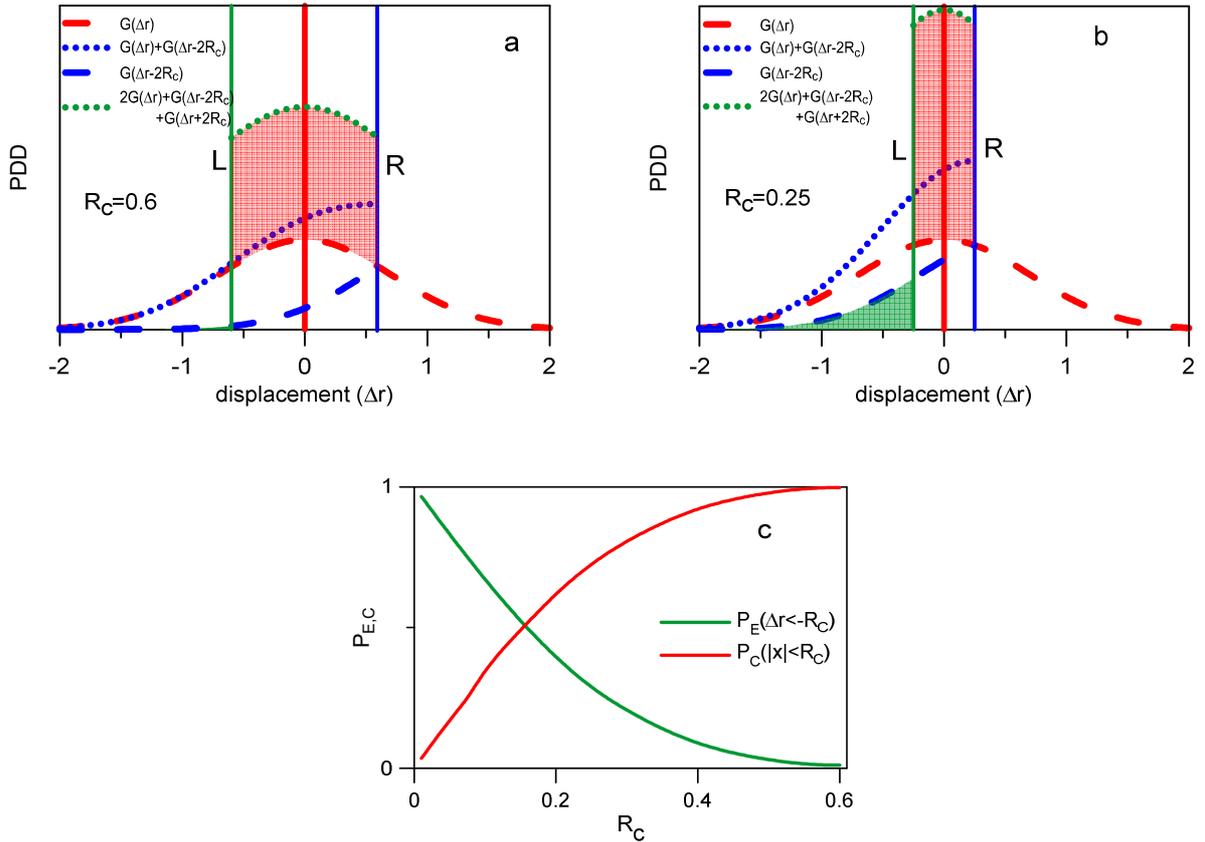

FIG. 2. In (a) and (b) $G(\Delta r)=\exp(-\Delta r^2)/\sqrt{\pi}$ is the primary Gaussian PDD centred on $\Delta r=0$ (red dashed line). The distances of the vertical lines L (green) and R (blue) from the central (red) line represent gaps of width $R_c$ between particle surfaces. $G(\Delta r)+G(\Delta r-2R_c)$ (blue dotted line) is the resulting PDD modified by the presence of a particle surface (reflecting barrier) R. $G(\Delta r-2R_c)$ (blue dashed line) is the reflection, or bias, in the primary PDD, effected by this barrier of which the green shaded part ($\Delta r<-R_c$) is transmitted to and beyond (the surface of) particle L. This is negligible for $R_c$=0.6 in (a) but more pronounced for $R_c$=0.25 in (b). Reflection of $G(\Delta r)+G(\Delta r-2R_c)$ from surface L gives $2G(\Delta r)+G(\Delta r-2R_c)+G(\Delta r+2R_c)$ (green dotted line). The net change to the primary PDD in the presence of both surfaces, based on just one reflection from each surface, is given by the red

shaded area. (c) Red and green lines express the red and green shaded areas in (a) and (b); respectively the probabilities, $P_C(|\Delta r|<R_c)$ and $P_E(\Delta r<-R_c)$ (or its positive equivalent $P_E(\Delta r>R_c)$) of caging (C) and escape (E).

In the limit $\phi \to 0$ the PDD at any in time is expressed by the Gaussian, $G(\Delta r)=\exp(-\Delta r^2)/\sqrt{\pi}$. At finite $\phi$, a particle's movement, about $\Delta r=0$, will on occasion be blocked and reversed by a neighbouring particle. As illustrated in Fig. 2a and 2b, the effect of this can be expressed by the truncation and reflection of that part of the particle's central, putative symmetrical Gaussian PDD, $G(\Delta r)$, that is momentarily cut off by the presence of a nearest neighbour's surface at R[38]. The reflection, or negative bias, $G(\Delta r-2R_C)$, (green shaded area in Fig. 2a and 2b) extends to the left of the surface of a second neighbour L. The presence of particle L presents two possibilities; One is, that surface L reflects the biased PDD, $G(\Delta r)+G(\Delta r-2R_C)$, resulting in the distribution, $2G(\Delta r)+G(\Delta r-2R_C)+G(\Delta r+2R_C)$, of a particle confined between L and R. The difference between the latter and the original (unbounded) PDD is indicated by the red shaded area. With this possibility, confinement the particle's displacements to the neighbour cage, $\Delta r<|R_c|$, we identify mechanism C (caging). The second possibility is that L moves to the left to accommodate the bias. Here we identify mechanism E (escape) as that where displacements by the central particle beyond the cage, $\Delta r<-R_c$, are sampled. Put another way, mechanism E requires an accommodating cage fluctuation in the negative direction; displacement of particle L which, in turn may require the same of a particle, or particles to its left. In any case this involves cooperation of three or more particles. Although particle L effects an equivalent positive bias, for clarity of illustration only the negative bias is shown in Fig. 2.

The magnitudes of the red and green areas, $P_C$ and $P_E$ shown in Fig. 2(c), give the respective probabilities of these mechanisms. It is evident that the probability of mechanism E increases with $\phi$, ie, as the average gap, $R_c$, between particle surfaces decreases; the more closely spaced the particles the greater the degree to which their movements in exploring configuration space are likely to be in phase. On average the negative bias is balanced by the positive bias. However, as we show below, the issue is to what extent the bias in one direction persists before it is balanced by that in opposite direction.

It has been shown that the non-Gaussian parameter, $\alpha$, of a diffusing particle confined to a sphere is negative[39]. One can see more easily from the schematic in Fig. 2 that caging tends to clip the "wings" of the PDD; decrease the 4th and higher order moments relative to the second. It follows from the definition (Eq. (8)) that this gives negative values of $\alpha$. Conversely, the same parameter for mechanism E, which tends to enhance the 4th and higher order moments, is positive.

The preceding schematic considers a fixed PDD in one dimension and does not actually describe diffusing or otherwise moving particles. However, despite this limitation and simplicity of this one-dimensional schematic, it serves to demonstrate that, except when $<\Delta r^2> << R_c$, "escape" (E) through correlated cage fluctuations competes with cage confinement (C). Both mechanisms cause stretching of the MSD. Which dominates is indicated by the sign of $\alpha$. These basic inferences apply irrespective of a system's dimensionality.

From now we consider systems in three dimensions as generally studied by computer simulation and experiment in order to determine how mechanisms C and E can be read and distinguished in the time correlation functions defined in Sec. II.A, particularly the CAF and non-Gaussian parameter. First of all we note that the latter has been determined in numerous studies over the past 50 years, from some of the earliest MD simulations [12], various theoretical approaches to Brownian systems [40, 41], subsequent Brownian dynamics simulations [42, 43] and experiments [44, 45] and a host of more recent simulations [13, 46] and experiments [20] in the supercooled region. A few illustrations will be shown below. In all these, in their respective computational or experimental time windows, $\alpha(\tau)$ increases from zero to a maximum and decreases to zero at large $\tau$. It appears therefore that delays in the decays of time correlation functions result primarily, although not necessarily exclusively, from correlated movements of three or more particles, be that mechanism E or, in the MD results, irreversible activated processes.

To explore why the cage effect appears to be so benign we proceed by noting that the MSD of the putative PDD of Brownian particles initially increases as $<\Delta r^2(\tau)>\sim\tau$ (Fig. 1a). Then according to Eq. (9), $<\Delta r^{2n}(\tau)>\sim\tau^n$, each successive moment increases by factor $\tau$ faster than the previous moment. Thus, the wings – the higher order moments – of the PDD of a given particle are the first to be affected, truncated and reflected in the manner displayed in Fig. 2, by the proximity of another particle. So, accommodation of the resulting (negative) bias, which spreads as $\sim\tau^n$ ($n\geq 2$), by neighbouring particles is probabilistically more efficient than the diffusive spread ($\sim\tau$) of the putative PDD. In other words, configuration space is explored more efficiently by means of correlated cage fluctuations, along the direction, $\hat{q}$, of the scattering vector, than by means of the intrinsic isotropic diffusion. Moreover, the efficiency of this mechanism increases with packing fraction.

In this mechanism, without any other excluded volume effects, a particle's displacements accommodate and are accommodated by the displacements of its neighbours; there is no structural impediment such as caging to this accommodation. In this regard the dynamics of all particles (in the ensemble) are statistically equivalent as are their PDDs; The dynamics are homogeneous in space and time and probability (the PDD) is conserved locally. And to comply with the latter, growth of the higher ($n\geq 2$) order moments must be compensated by shrinkage of the MSD; Accordingly, stretching or delay in the growth of the MSD, expressed by stretching index $\gamma$ (see Fig. 1b), is a consequence of probability conservation rather than caging per se. Of course with increasing delay time memory of the asymmetry of the cage fluctuations is lost – the system equilibrates – and all moments of the PDD ultimately grow in a manner consistent with a Fickian, Gaussian process. This means that at some delay time the non-Gaussian parameter, $\alpha(\tau)$, has a maximum.

To recapitulate, mechanism E defined by the transmission of the correlation of cage fluctuations – the directed bias of a particle's PDD – presents a more efficient means of establishing equilibrium than the putative Brownian motion. So, any simpler spontaneous, non-cooperative process of particles changing their neighbours is inconsistent with the more efficient cooperative approach to equilibrium. Since mechanism E is inextricably tied to local conservation of probability, all moments of the PDD are slaved to each other. Thus, among the signatures in experimental and computational data by which this mechanism can be identified are (i) that the non-diffusive aspect of structural relaxation is a single process, common to all particles, that is "faster" than isotropic diffusion; (ii) that the maximum in the non-Gaussian parameter and maximum stretching of the MSD occur at the same delay time.

Another possibility is that, due to some yet unspecified structural impediment, *some* particles, somewhere in this sea of cooperating particles, are caged at least for some time intervals in the experimental time window; some are subject to mechanism C and the rest move about cooperatively according mechanism E. As reasoned above, the most efficient way for a particle to escape its neighbour cage is by collective accommodating displacements of those neighbours – through a suitably large cage fluctuation. However, in order to conserve probability, at least globally – to conserve the number fraction of caged particles – such escape or transition, C→E, of a caged particle to the ranks of cooperating particles in one location requires a delayed compensatory transition, E→C, by another particle in another location. This suggest that caging generates another type of fluctuation, C⇆E, which is non-local in time and space. One anticipates such fluctuations to be manifested by a delay in the decay of time correlation functions that is intrinsically dependent on q. In addition, they give a delayed positive contribution to $\alpha(\tau)$. Accordingly, one expects the maximum in $\alpha(\tau)$ to be delayed relative to the maximum stretching of the MSD.

Of course caging does not prevent equilibrium from being attained to within some specified accuracy at sufficiently large delay time, whence memory of caging is lost and associated fluctuations, C⇆E, have decayed. However, the attendant inefficiency in exploring configuration space, relative to that of mechanism E, does affect the manner or rate by which this limit is approached. In addition it's not obvious how stretching of the MSD due to caging can be discerned from stretching due to mechanism E. So, rather than exploring the approach to equilibrium via the MSD and the usual time correlation functions of the particle number density (Eq.(1)), for example, the presence or absence of caging should be more directly exposed in the more subtle, higher order dynamical properties such as the non-Gaussian parameter and time correlation functions of the particle currents (Eq. (3) and (5)).

It is perhaps appropriate to recall that we are still discussing a system of Brownian particles; undamped phonons are precluded. However, the cage escapes in these systems, viewed locally, may appear similar to phonon activated hops and give similar contributions non-Gaussian contributions to the particles' displacement statistics[22]. However, the two escape mechanisms obviously differ fundamentally and have different long term consequences; C⇆E fluctuations being reversible and phonon activated processes being irreversible. We return to the role of (undamped) phonons in Sec. III.B.

## III. DISCUSSION

### A. Identification of mechanisms E and C⇌E in experimental data.

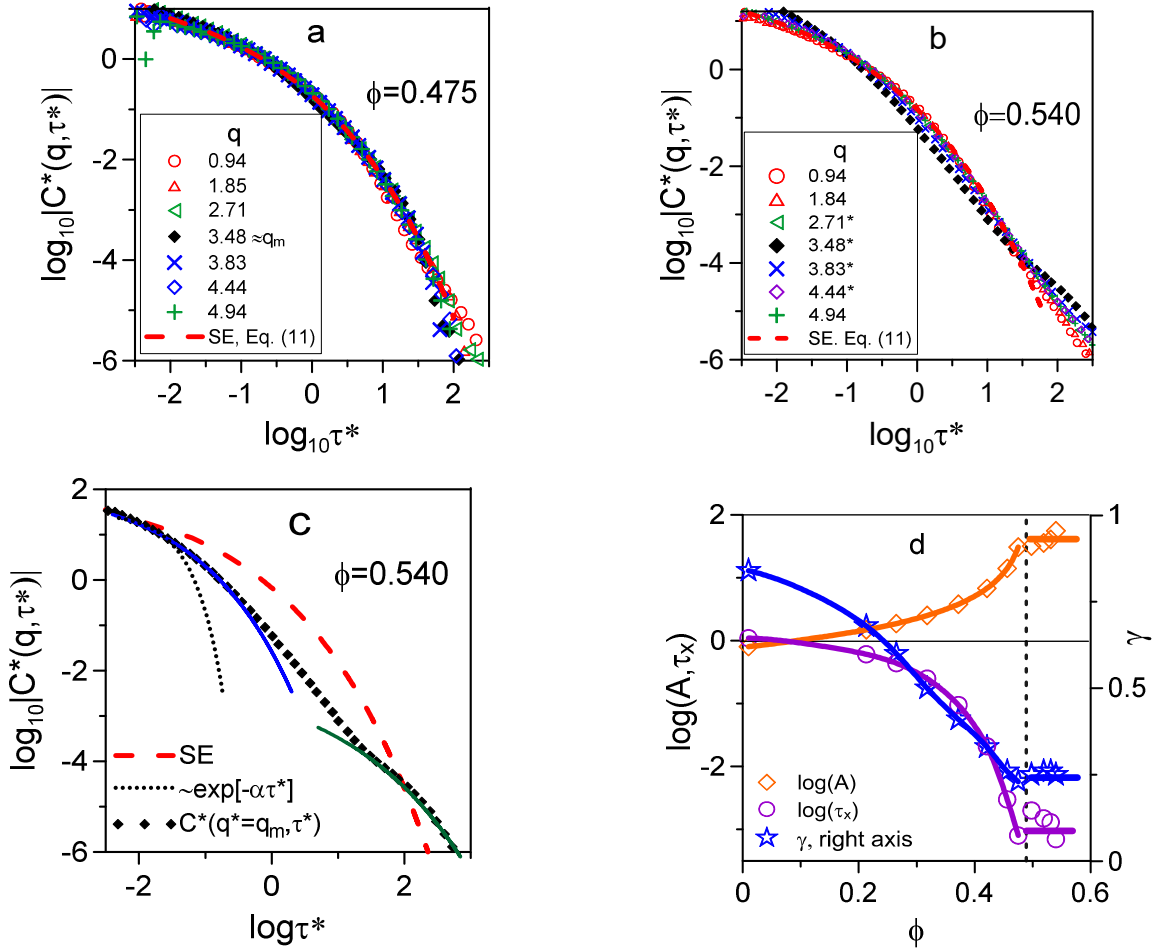

FIG. 3. Double logarithm plots of the SCAFs, $C^*(q,\tau^*)$ (Eq. (5)), at packing fractions (a) $\phi=0.475<\phi_f$ and (b) $\phi=0.540>\phi_f$ and spatial frequencies, q, indicated (Note that q is expressed in units of 1/a). In (b) variations in concavity, beyond experimental noise, are found in the decay of $C^*(q=q^*,\tau^*)$ for those values of q marked with an asterisk (*). (c) Illustrations of $C^*(q_m^*,\tau^*)$ and the stretched exponential (SE) (Eq. (11) below) that approximates $C^*(q\neq q^*,\tau^*)$. The latter has been translated along the y-axis so that it coincides with $C^*(q_m^*,\tau^*)$ at the left extremity of the time window. Continuous lines are drawn to indicate the presence, in $C^*(q_m^*,\tau^*)$, of processes faster (blue line) and slower (green line) than the SE≅$C^*(q\neq q^*,\tau^*)$. (d) Parameters A, $\tau_x$ and $\gamma$ of the SE fits to $C^*(q,\tau^*)$ for $\phi<\phi_f$ (to the left of vertical dashed line located at $\phi_f=0.494$) and fits to $C(q\neq q^*,\tau^*)$ for $\phi>\phi_f$. Lines are drawn as guide through average results. See discussion for further explanation. See Ref. [35, 36] for the source of some of these and additional data.

We consider a few results that are typical of those obtained from previous DLS experiments on suspensions of hard spheres[35, 36]. Fig. 3a and 3b show SCAFs, $C^*(q,\tau^*)$, defined in Eq. (5). The double-logarithm of the absolute values are plotted to capture the wide dynamic range and because $C^*(q,\tau^*)$ is negative in the experimental time window. In this window then, $C^*(q,\tau^*)$ exposes the correlation of *reversals* in the particle current along the direction $\hat{q}$. Note, in particular in Fig. 3a, that $C^*(q,\tau^*)$ shows no systematic dependence on spatial frequency, q, over approximately 6 decades of decay. For the larger packing fraction (Fig. 3b) there are deviations from this scaling in the form of subtle variations in concavity that, for those spatial frequencies labelled q*, cannot be encompassed by experimental noise. Fig. 3c attempts to distinguish more clearly the two types of decay seen for $C^*(q \neq q^*, \tau^*)$ and $C^*(q=q^*, \tau^*)$. We discuss this more fully below.

We turn first to the result, in Fig. 3a, for the thermodynamically stable example ($\phi < \phi_f$). The change of variable, $\tau^* = q^2 D(q)\tau$, in this case has effectively removed explicit dependence on $q$[35, 36]. This result is significant and pivotal to the remaining discussion. It implies that all interactions among the suspended particles, which for hard spheres constitute just the hydrodynamic interactions, are captured by the short-time diffusion coefficient, $D(q)$. In other words, hydrodynamic interactions are already established on the experimental time scale ($>10^{-6}$s.). This is expected from other works[47-49] which show rather more strongly that hydrodynamic interactions are established in the time ($\sim 10^{-12}$s.) for sound to propagate between the particles. Recall from Sec. II.A that the coefficient $D(q)=H(q)/S(q)$, where $H(q)$ accounts for hydrodynamic interactions, characterises the initial diffusive structural decay of a system of Brownian particles[4, 50]. This process is isotropic and its components in the direction, $\hat{\mathbf{q}}$, mirror symmetrical. Rescaling the delay time in terms of that, $1/D(q)q^2$, that characterises this diffusive process exposes, in $C^*(q,\tau^*)$, any correlation of fluctuations along this direction (see Eq. (4)). More specifically, any deviation of $C^*(q,\tau^*)$ from an exponential function of $\tau^*$ exposes a delay in structural relaxation due entirely to the correlation of reversals along the direction, $\hat{\mathbf{q}}$ (see Eq. (4)). So while the dynamical effect of the structure, $S(q)$, being entirely encapsulated in $D(q)$, is to *slow* the diffusive relaxation of that structure, through cage fluctuations, by a factor, $D(q)/D_0$, *it does not delay* that relaxation process; that delay is due the correlation of cage fluctuations.

We digress briefly to clarify the terminology. In atomic fluids the CAF, or its frequency spectrum, exposes, aside from the heat mode, acoustic excitations – longitudinal phonons – whose frequency and damping vary with $q$[33, 51]. It has also been established for many cases that around the fluid's freezing point its sound dispersion curve has a propagation gap – a window of spatial frequencies where longitudinal phonons are overdamped[52, 53]. We return to this property of liquids below. For the Brownian systems all motion is overdamped by definition and, in analogy with the terminology used in molecular fluids, we refer to the collective modes exposed by the SCAF as "overdamped phonons".

Continuing with the discussion of Fig. 3a, we see that the SCAF can be approximated by a stretched exponential (SE),

$$C^*(q,\tau^*)=A \exp[-(\tau^*/\tau_x)^\gamma]. \quad (11)$$

Although the SE is not the only function by which these results can be approximated, its parameters, unlike those of a polynomial for instance, permit physical interpretation. Accordingly, we characterise the overdamped phonons with an average amplitude, A, and a distribution of decay times whose average is $\tau_t = \tau_x \Gamma(1/\gamma)/\gamma$ ($\Gamma(y)$ is the Gamma function)[54]. The fitting parameters A, $\tau^*$ and $\gamma$ are shown in Fig. 3d, where we see that A increases with $\phi$, and that the characteristic/average decay time is less than one ($\tau_t \sim \tau_x < 1$) and decreases rapidly with $\phi$. The implication is that with increasing packing fraction the amplitude of the overdamped phonons increases and they are, on average, transmitted at an increasing rate ($\sim 1/\tau_x$). These characteristics of $C^*(q,\tau^*)$ are consistent with those required the most efficient exploration of configuration space by mechanism E introduced in Sec. II.B.

In brief, the independence of $C^*(q,\tau^*)$ of q confirms that the particles' motions are overdamped (Brownian). So, the collective consequences, mechanism E, are due entirely to excluded volume effects. As far as we can ascertain from the existing data[35, 36], this applies to thermodynamically stable suspensions (ie, $\phi<\phi_f$).

In contrast to the preceding, the scaling fails for the sample known to be in the metastable region ($\phi>\phi_f$) (Fig. 3b). As illustrated in Fig. 3c, in this case we can identify two different types of decay;

(i) For spatial frequencies, $q^*$, around the location, $q_m$, of the structure factor maximum there is a variation in concavity in the decay of $C^*(q,\tau^*)$ (black diamonds) which bridges a process faster (blue line) than the SE (red dashed line) that describes mechanism E and one that, for the case shown, is at least one decade slower (green line). Clearly, both processes are strongly delayed – stretched – with respect to an exponential decay (black dotted line).

(ii) In the other type, found for the remaining spatial frequencies, $q \neq q^*$, $C^*(q,\tau^*)$ has the signature of mechanism E insofar as it can be approximated by a SE that shows no systematic variation with q ($\neq q^*$). Moreover, within the noise of the available data [35, 36] the parameters of the SE fitted to the $C^*(q \neq q^*, \tau^*; \phi>\phi_f)$ are approximately the same as those of the SE fitted to $C^*(q \neq q^*, \tau^*; \phi \approx \phi_f)$ (see Fig. 3d). In other words, the collective dynamics for spatial frequencies $q \neq q^*$ corresponds to mechanism E of the stable suspension at $\phi \approx \phi_f$ and between $\phi_f$ and $\phi_g$ show no systematic variation with $\phi$.

The purpose of Fig. 4 is to quantify these variations in concavity with the quantity $L(q^*,\tau^*)=d\log C^*(q^*,\tau^*)/d\log\tau^*$ shown in the left panel. Obviously, a further numerical derivative of the data introduces further errors. These are indicated in the in the amplitude, $\Delta L(q^*)$, and width, $\Delta\tau^*$, of the oscillation in $L(q^*,\tau^*)$ (Fig. 4b), the quantities indicative of the magnitude and delay between the two processes of the C⇋E fluctuations identified under (i) above. While quantitatively these results must be treated with caution they, nevertheless, provide trends in the more subtle aspects of the structural relaxation in the metastable system. One is that the oscillation first appears when the packing fraction exceeds $\phi_f$ for spatial frequencies $q^*$ in close proximity to $q_m$. Another is that with increasing $\phi$ the amplitude of the oscillation intensifies and spreads to other q, in both directions; there is a widening of the spatial gap, $q_m-\delta q \lesssim q^* \lesssim q_m+\delta q$, from which overdamped phonons are excluded. A similar conclusion was reached by a different approach, by combining measurements of the coherent ISF (Eq. (1)) and self ISF (Eq. (6))[55].

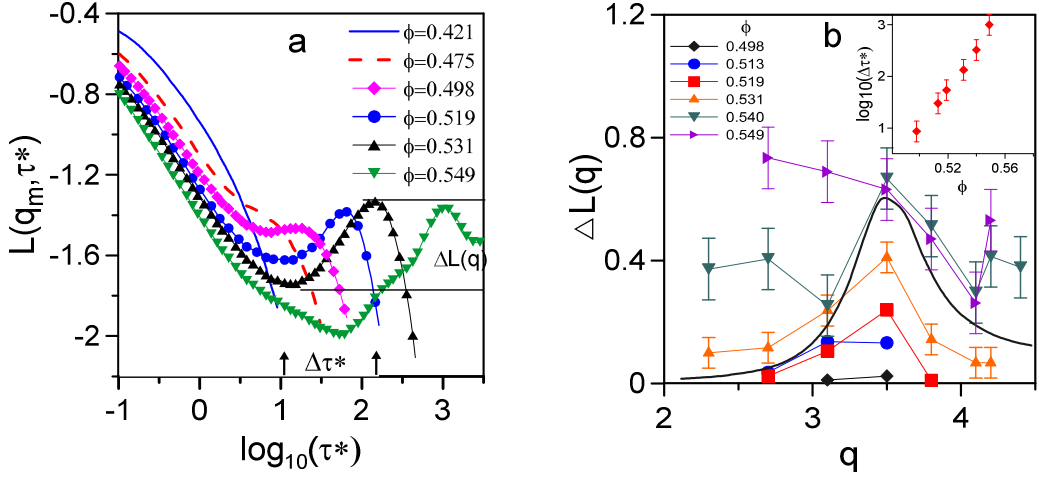

FIG. 4. (a) $L(q_m,\tau^*)=d\log C^*(q_m,\tau^*)/d\log\tau^*$ for values $\phi$ indicated. Black arrows and lines illustrate, for $\phi=0.531$, differences in time, $\Delta\tau^*$, and magnitude, $\Delta L(q)$, between extrema in $L(q_m,\tau^*)$. (b) Main figure; $\Delta L(q)$ versus $q$ for volume fractions indicated. The solid black line is the $S(q)/5$ at $\phi=0.5$. Inset; $\Delta\tau^*$ at $q_m$ (red diamonds). See Ref. [35, 36] for these and additional results.

In summary, in the metastable system there is a reciprocal space separation of the collective dynamics into fluctuations C⇆E, for $q=q^*$, and E, for $q\neq q^*$. To the extent that C⇆E fluctuations vary with $q$ they are non-local.

We now turn to consider relevant MD results for hard spheres. Being consequences of excluded volume effects, mechanisms E and C⇆E apply equally to fluids of particles subject to Newtonian dynamics. However, the associated signatures by which we have identified these mechanisms in suspensions can only be exposed unambiguously under conditions where sound is overdamped. As mentioned in Sec. I, gaps in sound dispersion curves – windows of spatial frequency where sound is overdamped – appear quite generally, it seems, for molecular fluids at sufficiently large density or low temperature[52, 56, 57]. For the hard sphere system this gap, centred on $q_m$, first emerges at $\phi\approx 0.45$ and disappears at $\phi\approx\phi_f$[58]. Aside from being evident directly in sound dispersion curves, the occurrence of overdamping of sound is also evident in the manner of decay of the negative long time tail of the CAF[58]. Where, for $0.45<\phi<\phi_f$ and $q\approx q_m$, sound is overdamped these decays can be approximated by a SE of which the characteristic decay time decreases with $\phi$. These are the characteristics of the overdamped phonons of mechanism E found for the colloidal system above. Then for $\phi>\phi_f$ these slow negative decays develop the variations in concavity symptomatic of fluctuations C⇆E. So, also in the fluid of particles subject to Newtonian motion we can identify the onset of caging, or suppression of overdamped phonons, around $\phi_f$, albeit by just a single spatial frequency $q^*\approx q_m$. Whether, with increasing $\phi$, the gap in which overdamped phonons are then excluded spreads in the same manner as that found for a suspension remains to be determined.

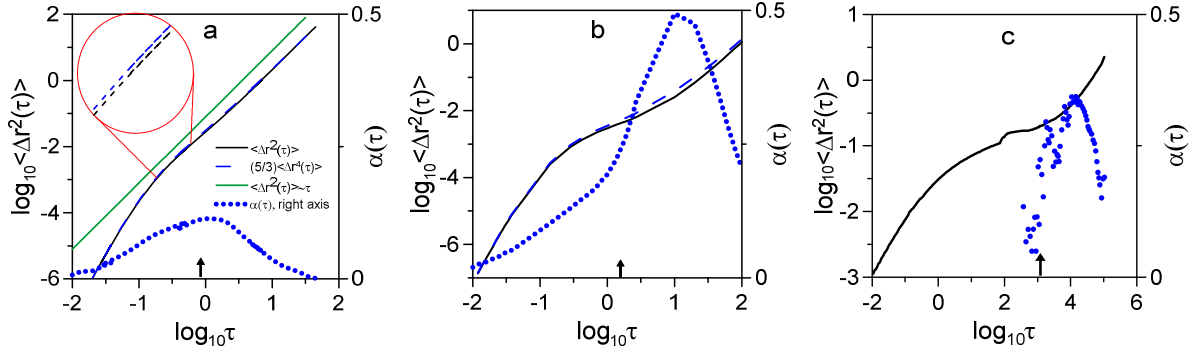

FIG. 5. $\log_{10}\langle\Delta r^2(\tau)\rangle$, black lines; $\log_{10}[(5/3)\langle\Delta r^4(\tau)\rangle]$, blue dashed lines; $\alpha(\tau)$ blue dots (right axis); (a) $\phi=0.481$ ($<\phi_f$), and (b) $\phi=0.547$ ($>\phi_f$) are MD results. In (a) the green line indicates diffusion, $\langle\Delta r^2(\tau)\rangle\sim\tau$. The arrows indicate delay times, $\tau_m$, where the MSD is maximally stretched. (c) DLS results for $\phi=0.560$

As implied in Sec. II.B the location of the maximum of the non-Gaussian parameter relative to the maximum stretching of the MSD is another signature of the presence of the fluctuations, C⇆E, indicative of caging. This parameter is a subtle property and experimental determination is not trivial and subject to error. Fortunately, due to a great deal of interest in glass transition dynamics over recent decades there are ample accurate results from computer simulations and experiment for metastable, or supercooled, fluids [13, 20, 59]. All of these indicate that the maximum in $\alpha(\tau)$ is delayed with respect to $\tau_m$. Fig. 5 presents illustrative results of the time trace of the MSD and the non-Gaussian parameter determined from MD and DLS on hard sphere systems. Within the uncertainties the delay is absent in the stable state (Fig. 5.a), while in the metastable state (Fig. 5b and 5c) the delay is about one decade in time. As proposed in Sec. II.B, it is this delay, rather than the non-Gaussian nature of the PDD as such, that provides the evidence for the non-local, heterogeneous nature of mechanism C⇆E.

Less attention in this respect has been devoted to fluids of more moderate density. So the evidence for mechanism E is less compelling. Nonetheless, despite the subtlety of the stretching one can see in Fig. 5a that, within computational noise, the maximum in $\alpha(\tau)$ occurs at $\tau_m$.

## B. Consequences of caging.

As already pointed out, the scaling of the delay time in terms of $1/D(q)q^2$ that renders $C^*(q,\tau^*)$ independent of q can apply only when *all* interactions among the particles are captured by the short diffusion coefficient. This requires that hydrodynamic interactions are the only interactions and that these are established on the experimental time scale; i.e., the scaling (seen in Fig. 3a) can apply only for a system of Brownian of hard spheres. Accordingly, as first pointed out in Ref. [35], failure of the scaling by way of variations in concavity, for $q=q^*$ (Fig. 3(b)), indicates that memory of hydrodynamic interactions persists on the experimental time scale; the particles' motions are no longer Brownian. The implication is that our discussion of the collective, but reversible, dynamics of the metastable suspension in terms of E and C⇆E fluctuations alone is incomplete.

At this stage it's appropriate to mention two points. First, the quantity $D(q)q^2$, determined by fitting an exponential function of delay time to the initial decay of the ISF (see Sec. II.A), is the simplest approximation that quantifies the decay of the fastest process detected by DLS.

This does not imply that this process is necessarily diffusive. The validity of that implication has been established here only for a suspension of hard spheres for $\phi<\phi_f$. By the same token, for $\phi>\phi_f$ and $q=q^*$, $D(q)q^2$ is merely a fitting parameter whose physical significance is unclear.

Second, it goes without saying that deviations from scaling, as expected in the presence of finite range interactions, are not in conflict with Brownian motion per se, so long as $C^*(q,\tau^*)$ is of a form, such as a SE (Eq. (11)), consistent with a distribution of overdamped modes. The conflict arises when such deviations are of a form, such as non-monotonicity in the derivative of the CAF seen in Fig. 4a, that is inconsistent with a distribution of exponential functions of delay time.

The incompleteness just mentioned may be exposed more starkly, albeit in light of the uncertainties somewhat speculatively, by considering the spatial gap, $q_m-\delta q \lesssim q^* \lesssim q_m+\delta q$, from which overdamped phonons are excluded. Should this gap spread more or less symmetrically about $q_m$, as suggested by the trends in Fig. 4b and Ref.[55], then beyond some packing fraction, $\phi_g$, cage fluctuations would be correlated only for spatial frequencies in excess of around $2q_m$, or length scales $\pi/q_m$; cage fluctuations would be localised to the nearest neighbours. All particles would then effectively be caged and the system would have no means to equilibrate.

The suggested dynamical termination at $\phi_g$ ignores the persistence of the memory of hydrodynamic interactions. This ultimately stems from the fact that caging or geometrical confinement of particles impairs their ability to respond, or recoil, in collisions and dissipate their thermal energy. While these momentum sinks or, in the language of Sec. 1, "energy hotspots", persist they can activate irreversible, ergodicity restoring processes that may ultimately lead to nucleation of the crystal phase. Still, in view of the works [22-24] mentioned in Sec. I and the results discussed above, it's evidently possible to construct simple systems, usually with a small spread in particle radii[30], in which stationarity in their metastable state is maintained long enough to measure time correlation functions. A suspension of hard sphere particles is one system that exhibits a fair approximation to the ideal glass transition, at $\phi_g \approx 0.57$[27, 60]. However, this is also the packing fraction in excess of which irreversibility – aging – cannot be ignored[61].

## IV. CONCLUSIONS

We have studied a system of Brownian hard spheres with the aim of determining the collective dynamical consequences of packing effects among the particles. The two consequences identified here differ by whether or not any of the particles are caged by their neighbours.

In the absence of caging on the experimental time scale the Brownian movements of every particle accommodate and are accommodated by the movements of their neighbours. As such, all particles (in the ensemble) are dynamically equivalent and the system's dynamics are homogeneous. This mutual accommodation is realised through overdamped phonons; transmission of the correlation of cage fluctuations (referred to above as mechanism E). On average these phonons decay faster than the diffusive decay of the intrinsic cage fluctuations themselves and, as such, they are the more efficient means of establishing equilibrium; the more so as the packing fraction is increased. By the same token, they effect the delay in the decay (ie, stretching) of time correlation functions of the particle number or current density.

Caging of some particles triggers fluctuations (referred to above as C⇆E) that are non-local in time and space. The reason we have offered for this is, for probability to be conserved globally, the "escape" of a caged particle at one location by mechanism E must wait for another to become caged. These non-local fluctuations have been identified by a second, delayed step in the decay of the time correlation function of the longitudinal particle current and displacement in time of the maximum of the non-Gaussian parameter with respect to the maximum stretching of the MSD.

As far as resolution of the present experimental data allows us to infer, the collective dynamics of the suspension for packing fractions up to the freezing value, $\phi_f$ can be characterised by (longitudinal) overdamped phonons. On entering the metastable phase caging and the attendant C⇆E fluctuations emerge, and exclude overdamped phonons, for spatial frequencies around the position of the main peak in the structure factor. The spatial window from which overdamped phonons are excluded widens with increasing packing fraction.

Similar mechanisms have been identified, and used to distinguish stability from metastability, in MD simulations. Although in this case, of particles subject to Newtonian dynamics, such identification has been possible only for conditions, $\phi>0.45$ and $q \approx q_m$, where sound is overdamped.

These mechanisms have been identified in MD and DLS results only for systems of hard spheres. It would be of some interest to see if the above conclusions apply equally to systems of particles with finite range interactions.

## ACKNOWLEDGEMENTS

We thank Sylvia Bassett, Gary Bryant, Peter Daivis, Jacques Joosten, Martin Oettel and Peter Pusey for their valuable comments on this work